\documentclass[reprint,aps,prd,superscriptaddress,nofootinbib,floats,showpacs]{revtex4-2}

\usepackage{lmodern}

\usepackage[]{graphicx}
\usepackage[]{graphics}
\usepackage{amsmath}
\usepackage{amssymb}
\usepackage{tensor,mathtools, amsthm} 
\usepackage{epsf}                                                                                           
\usepackage{color}                   
\usepackage{verbatim}
\usepackage[mathscr]{euscript}
\usepackage{mathrsfs}

\usepackage{hyperref}

\usepackage{enumerate}
\usepackage[T1]{fontenc}

\usepackage{bbm}

\newtheorem{theorem}{Theorem}[section]
\newtheorem{definition}{Definition}[section]

\begin{document}
\author{Ruijue Yan}
\affiliation{School of Physics and Astronomy, Key Laboratory of Multiscale Spin Physics (Ministry of Education), Beijing Normal University, Beijing 100875, China}

\author{You Ding}
\affiliation{School of Science, Beijing Jiaotong University, Beijing, 100044, China}

\author{Yongge Ma}
\email{mayg@bnu.edu.cn}
\affiliation{School of Physics and Astronomy, Key Laboratory of Multiscale Spin Physics (Ministry of Education), Beijing Normal University, Beijing 100875, China}

\author{Cong Zhang}
\affiliation{School of Physics and Astronomy, Key Laboratory of Multiscale Spin Physics (Ministry of Education), Beijing Normal University, Beijing 100875, China}

\title{Dynamics of Regge calculus with torsion}

\date{\today}

\begin{abstract}
The discrete geometry on an $n$-dimensional simplicial manifold are studied, in order to incorporate torsion into Regge calculus. In each simplex, the edge vectors are assigned to the edges to encode the information of their lengths and directions. In addition, the holonomies along the curves across the interfaces of two adjacent simplices are represented by the internal gauge group elements. The torsion manifests itself as the difference between an edge vector on an interface belonging to one simplex and the parallel transported edge vector, via the holonomy, of the same edge but belonging to the adjacent simplex. The simplicial Einstein--Cartan actions are then constructed as functions of the edge vectors and holonomies in three and four dimensions with Euclidean and Lorentzian signatures, respectively. It is shown that they return to the corresponding Regge actions for the torsion-free cases. The variations of the 4-dimensional discrete action with respect to the edge vectors and holonomies, respectively, give two equations of motion. It is shown that the former is consistent with the corresponding equation in the continuum theory, and the torsion-free holonomies satisfy the latter equation as in the continuous case. Thus, the resulting theory on the simplicial manifold can be regarded as a discrete analogue of Einstein--Cartan theory.
\end{abstract}

\maketitle

\section{Introduction}

General relativity is conventionally formulated on a Riemannian manifold, where the metric tensor is the fundamental dynamical variable and the spacetime connection is assumed to be the unique torsion-free, metric-compatible Levi-Civita connection. Geometrically, however, the metric and the connection represent distinct structures: the former determines lengths and angles, while the latter governs parallel transport. Since metric compatibility and vanishing torsion are logically independent assumptions, there is no fundamental reason to impose the torsion-free condition \emph{a priori}. Relaxing this requirement naturally leads to the Riemann-Cartan geometry \cite{Trautman:2006fp, Hehl:1976kj}, in which the tetrad and spin connection are treated as independent variables.

The corresponding first-order formulation of gravity is called the Einstein--Cartan theory and, in vacuum case, given by the Palatini action \cite{Palatini:1919ffw}. The variations of the Palatini action with respect to the tetrad and spin connection respectively yield two sets of equations of motion, one of which enforces the vanishing of torsion in vacuum. Consequently, the Palatini formulation reproduces classical general relativity for pure gravity. When fermionic matter is included, however, torsion becomes algebraically related to the spin density of the fermion field and generally does not vanish \cite{Hehl:1976kj,Shapiro:2001rz}. This suggests that torsion may play a fundamental role in the coupling between spacetime geometry and matter.

The study of torsion is also motivated by attempts to formulate a quantum theory of gravity. Loop quantum gravity (LQG) is based on the connection dynamics which can be obtained from the generalized Palatini action, where the tetrad and spin connection constitute the independent fundamental variables \cite{Holst:1995pc,Ashtekar:2004eh,Han:2005km}. In this framework, the torsion-free condition arises dynamically from the vacuum field equations rather than being imposed at the kinematical level. Furthermore, Regge calculus \cite{Regge:1961px}, a simplicial discretization of general relativity, plays an important role in the semiclassical description of LQG. The twisted geometries reduce to Regge geometries once the shape-matching conditions are imposed \cite{Dittrich:2008va,Freidel:2010aq,Long:2024pxb}. Moreover, the large-spin asymptotics of the EPRL spin-foam model is governed by the Regge action \cite{Engle:2007uq,Engle:2007wy,Barrett:2009gg,Han:2011rf,Han:2011re,Han:2013gna,Anza:2014tea}. These observations suggest that a simplicial formulation of Einstein--Cartan gravity may provide a natural bridge between torsion and quantum gravity.

In Regge calculus \cite{Regge:1961px, Williams:1991cd, Barrett:2018ybl, Dittrich:2008ar}, the spacetime manifold of dimension $n$ is discretized by triangulating it into $n$-simplices. Within each simplex, spacetime is assumed to be flat. The geometry of such a simplicial manifold is determined by the length of the edges, which correspond to the metric in continuum manifold. Non-trivial parallel transport occurs only when a vector crosses an interface between adjacent simplices. For the original definition of the transport in Regge calculus, the tangential and normal components of the vector are preserved across the shared face. This parallel transport can be regarded as a discrete analogue of the parallel transport defined by the torsion-free Levi-Civita connection in continuum geometry. With this definition, one can parallelly transport a vector along a closed loop surrounding each hinge, i.e., the $(n-2)$-simplex where these $n$-simplices meet. While the vector is orthogonal to the hinge, the angle between the initial vector and the vector after parallel transport around a hinge is called the deficit angle. This indicates curvature concentrated on the hinge, with the magnitude given by the deficit angle via the 2-dimensional Gauss-Bonnet theorem. The action in Regge calculus, whose continuum limit gives Einstein-Hilbert action \cite{Friedberg:1984ma}, was written as a functional of the edge length, whose variation yields the discrete version of the Einstein equation \cite{Barrett:1985nc}.

To introduce torsion into Regge calculus, in geometry, one has to figure out the following two issues. (i) Where is the torsion localized in the discrete simplicial manifold? (ii) What quantities can effectively characterize the torsion in the discrete simplicial manifold? Regarding the first issue, there exist three distinct treatments in literature. First, the torsion is introduced as a constant tensor field within each simplex; Second, the torsion is introduced as Dirac $\delta$-function condensed on the interface; Third, torsion singularities appear as conical defects (dislocations) on the hinges. In the first treatment \cite{Drummond:1986hb}, the simplices are rendered non-flat, so that the parallel transport depends on the curve along which it is taken, not merely on the interface across which the curve passes. In the second treatment, the effect of torsion contributes an additional rotation compared to the torsion-free case, in the parallel transport of vectors across the interface \cite{Drummond:1986hb,Holm:1991jda}. In the third treatment, the effect of torsion is represented by the so-called Burgers vectors, and the behavior of torsion is analogous to curvature singularities on hinges corresponding to the disclinations represented by deficit angles \cite{Schmidt:2001vd, Pereira:2002ff}. Although an explicit definition of the Burgers vector was given in Ref.~\cite{Ozakin2014AffineDO} and its relationship with torsion was elucidated in Weitzenb{\"o}ck Manifold where Riemann curvature is zero, the precise relationship between the Burgers vector and torsion remains unclear in the regions with non-zero Riemann curvature. Since the torsion in the Riemann-Cartan manifold contributes an additional rotation to the parallel transport of a vector, to keep the simplices flat inside and the parallel transport coincident with its continuum limit, we adopt the second treatment in this paper, such that the torsion is condensed on the interfaces as Dirac $\delta$-functions. Hence the connection is condensed on the interfaces in Regge geometry. It should be noted that the conclusion in Ref.~\cite{Drummond:1986hb} that this kind of torsion does not contribute to the Einstein--Cartan action is incorrect, since the contribution of torsion to curvature on the hinges was lost in the calculation \cite{Gronwald:1995mb}. Besides, the variables used to represent the torsion in \cite{Drummond:1986hb} are not well-defined under the gauge transformations. To describe the torsion freedom, the holonomy independent of the edge length is employed in this paper to represent the parallel transport across the interface. As in the first-order continuous theory, the information of the tetrad and spin connection are encoded in the edge (1-simplex) vectors and the holonomies passing across the interfaces ($(n-1)$-simplices), respectively.

The remainder of this paper is organized as follows. In Sec.~\ref{sec:Geometry}, the discrete Riemann-Cartan geometry is constructed on a simplicial manifold. We take the edge vectors and the holonomies as basic variables, and formulate the curvature and torsion based on this geometry. The integral form of the torsion in the regions with zero Riemann curvature is essentially consistent with the Burgers vector defined in \cite{Ozakin2014AffineDO}. In Sec.~\ref{sec:3D-action}, with the edge vectors and the holonomies as the fundamental variables, the 3-dimensional simplicial Einstein--Cartan action with Euclidean signature is obtained by two different approaches that yield the same expression. This expression is different from the result obtained in Ref.~\cite{Caselle:1989hd}, in which the Regge action was reformulated in the first-order formalism. In Sec.~\ref{sec:reduce to tor-free}, it is shown that the action obtained in the last section can be reduced to the Regge action for torsion-free holonomies. In Sec.~\ref{sec:4D-expand}, the simplicial expression of the Einstein--Cartan action is generalized to the 4-dimensional Lorentzian case. In Sec.~\ref{sec:EoM}, the equations of motion are studied. It is shown that the equation of motion obtained by varying the edge vectors is consistent with the corresponding equation in the continuous Einstein--Cartan theory, similar to the results of \cite{Barrett:1994nn, Barrett:1985nc}. Moreover, the torsion-free holonomies satisfy the equation of motion obtained by varying the holonomies. Hence the resulting framework can be viewed as a discrete analogue of Einstein–Cartan theory. The results are summarized and discussed in Sec.~\ref{sec:conslusion}. In Appendix~\ref{sec:non-Abe Stokes theom}, we review the non-Abelian Stokes theorem and discuss the small-surface approximation, which provides the basis for constructing the discrete Einstein--Cartan actions. In Appendix~\ref{sec:Martix Log}, we clarify the definition of the principal matrix logarithm used throughout the paper and collect its properties required for the variation of the logarithm of a holonomy.

Throughout the paper, we will use $a,b,c,\dots$ to denote the spacetime indices, and $I,J,K,\dots$ to denote the internal indices.

\section{Geometry of simplicial manifold with torsion}\label{sec:Geometry}

To construct a discrete Riemann–Cartan geometry on a simplicial manifold, we employ the edge vectors and holonomies as fundamental variables. They are the discrete counterparts of the vielbeins and spin connections, respectively. Within this framework, the curvature is distributionally supported on $(n-2)$-dimensional hinges, while the torsion is characterized by the mismatch of the vielbeins across $(n-1)$-dimensional interfaces. We will also derive an integral representation of the torsion flux and show that it gives the Burgers vector associated with the simplicial geometry.

\subsection{Discrete variables and gauge symmetry}\label{ssec:Geometry l-h}

Let $\sigma=[v_0,\ldots,v_n]$ be an oriented $n$-simplex, which is the convex hull of the $n+1$ affinely independent vertices $v_0,\ldots,v_n$. Its $k$-dimensional faces are $k$-simplices spanned by $k+1$ of these vertices. In particular, $\sigma$ contains $\binom{n+1}{2}$ edges and $\binom{n+1}{3}$ triangular 2-faces, where $\binom{m}{k}$ denotes the binomial coefficient, i.e., the number of ways to choose $k$ elements from a set of $m$ elements, and the boundary of $\sigma$ consists of $n+1$ $(n-1)$-faces.

An internal vector $l_p^I$ is assigned to each oriented edge $p$ of $\sigma$, called an edge vector, with the internal index $I=1,\ldots,n$ in Euclidean signature and $I=0,\ldots,n-1$ in Lorentzian signature. Reversing the orientation of an edge reverses the corresponding edge vector. Thus, the component numbers contained in the edge-vector data of one $n$-simplex equal to
\begin{equation}
n\binom{n+1}{2}.
\end{equation}

The edge vectors are required to be compatible with the affine geometry of the simplex, by imposing the closure condition on each oriented triangular 2-face $f$. Explicitly, for the oriented boundary edges $p\subset \partial f$, the closure condition implies
\begin{equation}\label{eq:clo-condition l}
\sum_{p\in \partial f}\epsilon_{fp}l_p^I=0,
\end{equation}
where $\epsilon_{fp}=\pm1$ is the incidence number encoding the relative orientation between the edge $p$ and the oriented boundary $\partial f$ of $f$.

Although each of the $\binom{n+1}{3}$ triangular 2-faces has a closure equation, these equations are not all independent. To see this explicitly, let $v_0$ be a reference vertex and $p_{0\alpha}, \alpha=1,\ldots,n$, be the $n$ edges emanating from $v_0$. For each pair $(\alpha,\beta)$ with $\alpha<\beta$, the triangular 2-face $[v_0,v_\alpha,v_\beta]$ gives
\begin{equation}
l_{0\alpha}^I+l_{\alpha\beta}^I+l_{\beta0}^I=0 .
\end{equation}
Using $l_{\beta0}^I=-l_{0\beta}^I$, this implies
\begin{equation}\label{eq:edge-from-reference}
l_{\alpha\beta}^I=l_{0\beta}^I-l_{0\alpha}^I .
\end{equation}
Therefore, once the $n$ edge vectors $l_{0\alpha}^I$ are given, all remaining edge vectors are fixed by the closure equations on the triangular 2-faces containing $v_0$. There are precisely $\binom{n}{2}$ such triangular 2-faces. Moreover, the closure equations on triangular 2-faces not containing $v_0$ then follow automatically. For the triangular 2-face $[v_\alpha,v_\beta,v_\gamma]$, equation \eqref{eq:edge-from-reference} implies
\begin{equation}
l_{\alpha\beta}^I+l_{\beta\gamma}^I+l_{\gamma\alpha}^I
\!=\!
(l_{0\beta}^I-l_{0\alpha}^I)+(l_{0\gamma}^I-l_{0\beta}^I)+(l_{0\alpha}^I-l_{0\gamma}^I)
\!=\!
0 .
\end{equation}
Hence only $\binom{n}{2}$ triangular closure constraints are independent. The number of independent components carried by the edge vectors of one $n$-simplex is therefore
\begin{equation}
n\binom{n+1}{2}-n\binom{n}{2}= n^2 .
\end{equation}
Equivalently, the independent data are given by the $n$ edge vectors $l_{0\alpha}^I$, each of which has $n$ internal components.

The above counting has a direct geometric interpretation. The $n^2$ independent components of the edge vectors are precisely the number of components of a vielbein $e^I_a$ inside an $n$-simplex. More explicitly, by introducing local affine coordinates $x^a$ in the interior of $\sigma$, the $n$ edges emanating from $v_0$ give naturally the coordinate displacement vectors
\begin{equation}
l_\alpha^a=x^a(v_\alpha)-x^a(v_0),\qquad \alpha=1,\ldots,n .
\end{equation}
Since the vertices of $\sigma$ are affinely independent, the matrix $l_\alpha^a$ is invertible. A constant vielbein one-form $e^I=e^I_a dx^a$ inside $\sigma$ is then uniquely determined by requiring
\begin{equation}\label{eq:edge-vielbein-reference}
l_{0\alpha}^I=\int_{p_{0\alpha}}e^I_a dx^a=e^I_a l_\alpha^a .
\end{equation}
Thus the independent edge vectors reconstruct a unique constant vielbein on the flat interior of the simplex.

For a general edge $p_{\alpha\beta}$ oriented from $v_\alpha$ to $v_\beta$, the same vielbein gives
\begin{equation}
\int_{p_{\alpha\beta}}e^I_a dx^a=e^I_a\bigl[x^a(v_\beta)-x^a(v_\alpha)\bigr]=l_{0\beta}^I-l_{0\alpha}^I=l_{\alpha\beta}^I .
\end{equation}
Therefore, the triangular closure conditions ensure that the edge vector associated with an edge depends only on its endpoints. In this sense, the discrete edge-vector data may be viewed as the integral representation of a piecewise-constant vielbein field.

We now discuss the local gauge symmetry. Each $n$-simplex is equipped with its own local internal orthonormal frame. Hence a local gauge transformation is specified independently in each simplex by a group element
\begin{equation}
\Lambda(i)\in G,\, G=\begin{cases}
\mathrm{SO}(n), & \text{Euclidean signature},\\
\mathrm{SO}^+(1,n-1), & \text{Lorentzian signature},
\end{cases}
\end{equation}
where $i$ labels the simplex, and $\mathrm{SO}^+(1,n-1)$ denotes the proper Lorentz group in $n$ dimensions. Under such a gauge transformation, the edge vectors in the $i$-th simplex transform as
\begin{equation}\label{eq:Geo l gauge transformation}
\tilde l_p^I(i)=\Lambda(i)^I{}_J l_p^J(i).
\end{equation}
Equivalently, the corresponding constant vielbein transforms as
\begin{equation}
\tilde e^I_a(i)=\Lambda(i)^I{}_J e^J_a(i).
\end{equation}

Consider now two adjacent $n$-simplices $\sigma_i$ and $\sigma_j$ sharing a common $(n-1)$-face $\tau_{ij}=\sigma_i\cap\sigma_j$. The same geometric edge $p\subset\tau_{ij}$ should be described in both the local frame of $\sigma_i$ and the local frame of $\sigma_j$. We denote the corresponding edge vectors by $l_p^I(i)$ and $l_p^I(j)$, respectively. Consequently, on each edge there are several different vectors belonging to different simplices. Compatibility of the intrinsic geometry of the common interface means that the two vectors are related by a local gauge transformation
\begin{equation}\label{eq:interface-frame-matching}
l_p^I(j)=\Lambda_{ij}{}^I{}_J l_p^J(i),\qquad \Lambda_{ij}\in G .
\end{equation}
This relation expresses the change of local internal frame between the two descriptions of the same interface geometry. It should be distinguished from the holonomy introduced below, which represents a parallel transport by the discrete connection.

We now introduce the parallel transport of internal vectors across adjacent simplices. It is assumed that a parallel transport is trivial along any curve lying entirely inside a single simplex. Therefore, a non-trivial parallel transport can only arise when the curve crosses the common interface between two adjacent $n$-simplices. Let $\sigma_i$ and $\sigma_j$ be two adjacent $n$-simplices sharing a common $(n-1)$-face. For an oriented crossing from $\sigma_i$ to $\sigma_j$, the parallel transport is represented by a group element as
\begin{equation}
h_{ij}\in G,\, G=
\begin{cases}
\mathrm{SO}(n), & \text{Euclidean case},\\
\mathrm{SO}^+(1,n-1), & \text{Lorentzian case}.
\end{cases}
\end{equation}
Thus, if an internal vector $v^I$ is parallel transported from $\sigma_i$ to $\sigma_j$, it should satisfy
\begin{equation}\label{eq:Geo v parallel transport}
v^I(j)=h_{ij}{}^I{}_J v^J(i).
\end{equation}
Note that the group element $h_{ij}$ is assigned to the oriented crossing of the interface from $\sigma_i$ to $\sigma_j$, in the sense that it depends only on the interface crossing rather than on the detailed shape of the curve or the precise point at which the curve intersects the interface.

It should also be noted that the parallel-transport maps satisfy the following group-theoretic properties. First, reversing the orientation of the crossing gives the inverse transformation:
\begin{equation}
h_{ji}=h_{ij}^{-1}.
\end{equation}
Second, if a discrete path crosses successively from $\sigma_i$ to $\sigma_j$, and then from $\sigma_j$ to $\sigma_k$, the corresponding parallel transport is given by the product
\begin{equation}
h_{i\to j\to k}=h_{jk}h_{ij}.
\end{equation}
For a path crossing three consecutive interfaces,
$$\sigma_i\rightarrow \sigma_j\rightarrow \sigma_k\rightarrow \sigma_l ,$$
the total parallel transport is independent of how the products are grouped:
\begin{equation}
(h_{kl}h_{jk})h_{ij}=h_{kl}(h_{jk}h_{ij}).
\end{equation}
This associativity follows directly from the associativity of the group multiplication in $G$. Thus $h_{ij}$ represents the holonomy associated with the elementary interface crossing from $\sigma_i$ to $\sigma_j$. A general holonomy along a discrete path is obtained by ordered products of such elementary holonomies. In the simplicial setting, the spin connection is taken to be flat inside each simplex and distributionally supported on the interfaces between adjacent simplices. The discrete group element $h_{ij}$ may be regarded as the holonomy generated by such spin connection across the interface. Note that the holonomies should be compatible with the edge vectors, reflecting the compatibility between the spin connection and the coframe in Riemann–Cartan geometry.

It is straightforward to obtain the behavior of the holonomies under local gauge transformations. Consider a vector $v^I$ that is parallel transported from simplex $i$ to simplex $j$. Requiring that the parallel transport law remains valid after gauge transformation, i.e.,
\begin{equation}\label{eq:Geo til-v parallel transport}
\tilde{v}^I(j)=\tilde{h}_{ij}{}^I{}_J\tilde{v}^J(i),
\end{equation}
we obtain the gauge transformation law for the holonomy by using Eqs. \eqref{eq:Geo l gauge transformation} \eqref{eq:Geo v parallel transport} and \eqref{eq:Geo til-v parallel transport} as
\begin{equation}\label{eq:Geo holonomy gauge transformation}
\tilde{h}_{ij}=\Lambda(j)h_{ij}\Lambda^{-1}(i).
\end{equation}
Hence the gauge transformation of the holonomy is only related to the gauge transformations at the start and end points.

In summary, the edge vectors and holonomies provide a discrete description of the vielbein and spin connection, respectively, with an independent local $\mathrm{SO}(n)$ (or $\mathrm{SO}^+(1,n-1)$) gauge group acting on each simplex.

\subsection{Curvature and torsion on a simplicial manifold}

In a smooth manifold, the Riemann curvature 2-form is defined in terms of the spin connection 1-form $\omega^I{}_J$ by
\begin{equation}\label{eq:Geo Riem cur}
R^I{}_J:={\rm d}\omega^I{}_J+\omega^I{}_L\wedge\omega^L{}_J.
\end{equation}
In a piecewise-flat simplicial geometry described above, the curvature vanishes inside each $n$-simplex. Moreover, a single $(n-1)$-dimensional interface does not by itself carry a curvature defect. To see this, let us consider a closed curve starting from an arbitrary point $p$ inside a simplex, crossing the interface to an arbitrary point $q$ inside an adjacent simplex, and then returning to $p$ by crossing the same interface in the opposite direction. The holonomies associated with the two crossings are inverse to each other, and hence the total holonomy of this closed curve is the identity. Since this holds for arbitrary choices of $p$ and $q$, such loops detect no curvature on the interface. Therefore, the curvature can only be distributionally supported on the $(n-2)$-dimensional hinges, where several interfaces meet and a loop encircling the hinge can acquire a nontrivial holonomy. Thus only the curvature concentrated on the hinges contributes to the curvature part of the Einstein--Cartan action. This contribution will be calculated in the next section.

We now turn to the torsion. In a smooth manifold, the torsion 2-form is defined by
\begin{equation}
T^I:={\rm d}e^I+\omega^I{}_J\wedge e^J.
\end{equation}
Since torsion is a two-form, it is naturally characterized by its flux through a two-dimensional surface. We first derive a general integral expression for this flux in a curvature-free region, and then generalize it to a simplicial manifold.

Let $S$ be an oriented two-dimensional surface contained in a simply connected region with vanishing Riemann curvature. Choose a base point $x_0$ in this region, and let $h(x_0,x)\in G$ be the holonomy from $x_0$ to $x$ with $h(x_0,x_0)=\mathbbm{1}$. Because the curvature vanishes in this region, the parallel transport is path independent for a given endpoint. We define the covariantly transported vielbein by
\begin{equation}\label{eq:covariant e}
\mathscr{E}^I(x):=(h(x_0,x)^{-1})^I{}_J e^J(x) ,
\end{equation}
which satisfies
\begin{equation}
\begin{aligned}
{\rm d}\mathscr E^I=& {\rm d}\bigl((h^{-1})^I{}_J e^J\bigr)\\
=& (h^{-1})^I{}_J \left( {\rm d}e^J+\omega^J{}_K\wedge e^K \right)\\
=& (h^{-1})^I{}_J T^J .
\end{aligned}
\end{equation}
Therefore, the covariant torsion flux through $S$ is
\begin{equation}\label{eq:torsion flux stokes}
\int_S (h^{-1})^I{}_J T^J= \int_S {\rm d}\mathscr E^I= \oint_{\partial S}\mathscr E^I ,
\end{equation}
where the last equality follows from Stokes' theorem. This identity shows that, in a simply connected curvature-free region, the covariant torsion flux through a two-dimensional surface is given by the boundary integral of the covariantly transported vielbein. This quantity is the analogue of the Burgers vector associated with torsion \cite{Ozakin2014AffineDO}.

\begin{figure}[h]
\centering
\includegraphics[width=0.35\textwidth]{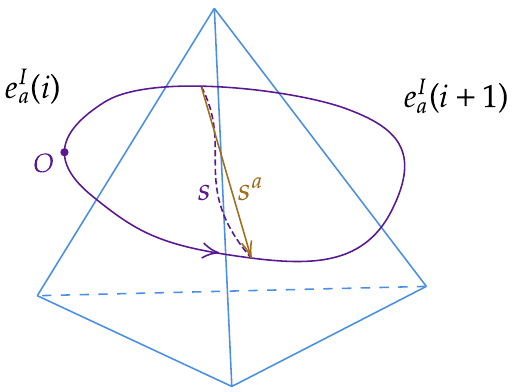}
\caption{Tetrahedral interface intersected by a 2-dimensional surface}
\label{fig:interface with 2D surface}
\end{figure}

Eq.~\eqref{eq:torsion flux stokes} can be generalized to a simplicial manifold. Consider a region in a 4-dimensional simplicial manifold consisting of two simplices sharing a common 3-dimensional interface. The Riemann curvature is zero inside this region, as discussed above. Given a 2-dimensional surface within this region that intersects the interface as a line $s$, we choose the starting point $O$ inside the $i$-th simplex. The loop integral of $\mathscr{E}^I$ along the boundary of this surface representing the 2-dimensional integral of the torsion inside the surface can be calculated as
\begin{equation}\label{eq:Geo torsion int simp}
\oint \mathscr{E}^I=s^ae^I_a(i)-s^a(h^{-1}_{ii+1})^I{}_Je^J_a(i+1),
\end{equation}
where $s^a$ is the displacement vector tangent to the interface, pointing from the starting point to the endpoint of the intersection line $s$, as shown in Fig.~\ref{fig:interface with 2D surface}. Eq.~\eqref{eq:Geo torsion int simp} implies that, on the simplicial manifold, torsion manifests itself as the mismatch between the projections of the vielbein onto the interface as seen from one simplex and that of the adjacent simplex after the parallel transport by the holonomy to the same local frame. Thus, the torsion-free situation corresponds to the condition
\begin{equation}\label{eq:torsion-free e-h}
\begin{gathered}
s^ae^I_a(i)=s^a(h^{-1}_{ii+1})^I{}_Je^J_a(i+1),\\
\forall s^a \text{ tangent to the interface},
\end{gathered}
\end{equation}
which implies that after parallel transport to the same local frame, the vielbeins induced from the two adjacent simplices agree on the common interface. Since the vielbein in each simplex has been defined by the edge vectors introduced above, Eq.~\eqref{eq:torsion-free e-h} indicates the matching condition for all edge vectors lying in the interface. Namely, for every edge vector $l_p^I$ tangent to the common interface, one has
\begin{equation}\label{eq:torsion-free l-h}
l_p^I(i)= (h^{-1}_{ii+1})^I{}_J l^J_p(i+1).
\end{equation}
Thus, the torsion-free condition requires that the edge vectors of the common interface, when expressed in the two local frames, are related precisely by the holonomy $h_{ii+1}$. Since the holonomy $h_{ij}$ belongs to either $\mathrm{SO}(n)$ in the Euclidean case or $\mathrm{SO}^+(1,n-1)$ in the Lorentzian case, the complement of the vielbein orthogonal to the $(n-1)$-dimensional interface also satisfies the same matching condition as in Eq.~\eqref{eq:torsion-free l-h} up to a sign, and this sign is fixed by the orientation condition. Hence no independent matching condition for the transverse frame direction is required.

Consequently, the holonomy satisfying the torsion-free condition above is precisely the torsion-free holonomy compatible with Regge geometry \cite{Regge:1961px}. This torsion-free holonomy condition will be used below to identify the torsion-free sector of the simplicial Einstein--Cartan action.

\section{Einstein--Cartan action in 3-dimensional Euclidean simplicial manifold}\label{sec:3D-action}

For simplicity, we first consider a 3-dimensional Euclidean manifold in this section. On the simplicial manifold, the Riemann curvature is condensed on the hinges,  so is the Einstein--Cartan action. Recall that the Einstein--Cartan action in a 3-dimensional differential manifold is given by
\begin{equation}\label{eq:3D-action 3D-EC action}
^3S_{EC}[e,\omega]=\int \frac{1}{2}\epsilon_{IJ}{}^Ke^I\wedge R^J{}_K,
\end{equation}
where $R^J{}_K$ is defined by Eq.~\eqref{eq:Geo Riem cur}, and throughout this paper, we adopt the convention $8\pi G=1$, with $G$ being the Newtonian gravitational constant. The fundamental variables here are the co-triad $e^I$ and the spin connection $\omega^I{}_J$. We need to express this action in terms of variables on the simplicial manifold (edge vectors and holonomies). As shown in the following subsections, the action integral on the hinges can be approximated in two approaches.

\subsection{Curvature average}\label{ssec:3D-action A}

In the first approach, we introduce a set of sufficient local conditions under which the Einstein–Cartan action can be factorized into an integral of the co-triad along one distinguished direction and an integral of the curvature over a transverse two-dimensional surface.

Given a vector field $b^a$ and a function $\zeta$ such that
\begin{equation}\label{eq:3D-action b-zeta}
b^a({\rm d}\zeta)_a=1,
\end{equation}
we consider a small 3-dimensional region and assume that the triad and $b^a$ satisfy
\begin{equation}\label{eq:3D-action triad gauge}
e_a^I(p)b^a(p)=b^I(z),
\end{equation}
where $p$ represents an arbitrary point in the foliation of $\zeta=z$ in the region. Moreover, we further assume in this region that the spin connection satisfies
\begin{subequations}\label{eq:3D-action omega req A}
\begin{align}
\label{eq:3D-action omega req1 A}
b^a\omega_a{}^I{}_J=0,\\
\label{eq:3D-action omega req2 A}
\mathcal{L}_{b}\omega_a{}^I{}_J=0,
\end{align}
\end{subequations}
where $\mathcal{L}$ denotes the Lie derivative. With the vector field $b^a$ and the function $\zeta$, a coordinate system can be introduced in the region such that $b^a=(\partial/\partial z)^a$ with right handed coordinates $\{x,y,z\}$. Eqs. \eqref{eq:3D-action omega req1 A} and \eqref{eq:3D-action omega req2 A} imply that the spin connection can be written in this coordinate system in the following form
\begin{equation}\label{eq:3D-action omega-xy}
\omega^I{}_J=\omega_x{}^I{}_J(x,y){\rm d}x+\omega_y{}^I{}_J(x,y){\rm d}y,
\end{equation}
and hence the Riemann curvature can be written as
\begin{equation}\label{eq:3D-action R-xy}
R^I{}_J=R_{xy}{}^I{}_J(x,y){\rm d}x\wedge{\rm d}y.
\end{equation}
Thus, in this region the Einstein--Cartan action can be decomposed as
\begin{equation}\label{eq:3D-action decomp int}
^3S=\frac{1}{2}\epsilon_{IJ}{}^K \int_l e_z^I {\rm d}z \iint_S R_{xy}{}^J{}_K {\rm d}x {\rm d}y,
\end{equation}
where
\begin{equation}
\int_l e_z^I {\rm d}z =\int_l b^I {\rm d}z =l^I.
\end{equation}
Using the non-Abelian Stokes theorem (see \eqref{eq:non-Abe int R-h} in Appendix \ref{sec:non-Abe Stokes theom}), the integral of the Riemann curvature can be approximated as
\begin{equation}\label{eq:3D-action intR=lnh}
\iint_S R^J{}_K\approx -({\rm Ln}h_C)^J{}_K,
\end{equation}
where $h_C$ is the holonomy along the curve $C=\partial S$ surrounding the 2-dimensional region $S$, with the orientation of $C$ chosen according to the right-hand rule with respect to $b^a=(\partial/\partial z)^a$. Hence $h_C$ belongs to the Lie group $\mathrm{SO}(3)$, and ${\rm Ln}h$ represents the logarithm of $h$ in matrix representation, which corresponds to the generator of $h$. It is well defined due to the compactness of $\mathrm{SO}(3)$. The details of the logarithmic calculation are provided in Appendix \ref{sec:Martix Log}. Thus, the action \eqref{eq:3D-action decomp int} can be rewritten as
\begin{equation}\label{eq:3D-action 3S dif to holonomy A}
^3S\approx -\frac{1}{2} \epsilon_{IJ}{}^K l^I({\rm Ln}h_C)^J{}_K.
\end{equation}
Here the holonomies come from the spin connection satisfying Eqs. \eqref{eq:3D-action omega req1 A} and \eqref{eq:3D-action omega req2 A}. Eq.~\eqref{eq:3D-action omega req1 A} implies that the holonomies along the integral curves of the vector field $b^a$ ($z$-axis) are all the identity. Let $C(t),t\in[0,1]$, be any curve in this region and $f:C\to C',C(t)\mapsto C'(t)$, be a continuous map such that $x(C(t))=x(C'(t)), y(C(t))=y(C'(t))$. Eq.~\eqref{eq:3D-action omega req2 A} implies that the holonomy along $C$ equals the holonomy along $C'$.

In the 3-dimensional simplicial manifold with torsion, we consider the region containing a hinge and choose the vector $b^a$ to be tangent to the hinge. Note that the 3-simplex is flat inside and the hinge belongs to all 3-simplices (tetrahedrons) surrounding it. Hence, we can give a set of parallel transport invariant fields $b^a$ inside each tetrahedron, and thus a vector field $b^a$ in this region. A consistent edge vector $l_p^I(i)$ can be assigned to the hinge by the following gauge transformations:
\begin{equation}
\Lambda^I{}_J(i)l_p^J(i)=\tilde{l}_p^I,
\end{equation}
where $p$ represents the edge as the hinge, and $i$ labels the $i$-th tetrahedron surrounding the hinge. Here the simplices can be labeled in a counterclockwise sequence $i=1,2,\dots,m$, with the orientation of this sequence chosen according to the right-hand rule with respect to the corresponding hinge $l^I$. Under this gauge, the triad and the $b^a$ field on the hinge satisfy
\begin{equation}
\tilde{e}^I_a(i)b^a(i)=\Lambda^I{}_J(i)e^J_a(i)b^a(i)=\Lambda^I{}_J(i)b^J(i)=\tilde{b}^I.
\end{equation}
With the same gauge transformation \eqref{eq:Geo holonomy gauge transformation}, the holonomy along a curve passing across an interface transforms as
\begin{equation}
\tilde{h}_{ii+1}{}^I{}_J=\Lambda^I{}_L(i+1)h_{ii+1}{}^L{}_K(\Lambda^{-1})^K{}_J(i).
\end{equation}
As the holonomy of a connection satisfying Eq.~\eqref{eq:3D-action omega req A} shares the same properties as the discrete holonomy discussed in Sec.~\ref{ssec:Geometry l-h}, the expression \eqref{eq:3D-action 3S dif to holonomy A} can be used to define the action of the discrete variables in this region. Moreover, since the holonomies along the closed curve around the hinge starting from points inside different simplices may differ from each other, we take the average of these holonomies to approximate the 2-dimensional integral of the Riemann curvature as
\begin{equation}
\iint_S R^J{}_K\approx -\frac{1}{m}\sum_{i=1}^m({\rm Ln}\tilde{h}_B^{(i)})^J{}_K,
\end{equation}
where
\begin{equation}
\tilde{h}_B^{(i)}=\Lambda(i)h_B^{(i)}\Lambda^{-1}(i),
\end{equation}
and
\begin{equation}
h_B^{(i)}\equiv h_{i-1i}h_{i-2i-1}\cdots h_{12}h_{m1}\cdots h_{i+1i+2}h_{ii+1}
\end{equation}
The loop defining $h_B^{(i)}$ follows the same orientation as the counterclockwise ordering of the simplices, namely the orientation determined by the right-hand rule with respect to the corresponding hinge $l_B^I$, consistently with Eq.~\eqref{eq:3D-action intR=lnh}.
Thus the action in this region is given by
\begin{equation}\label{eq:3D-action 3D simp action on hinge gauge}
^3S_B=-\frac{1}{2m_B}\sum_{i=1}^{m_B}\epsilon_{IJ}{}^K\tilde{l}_B^I({\rm Ln}\tilde{h}_B^{(i)})^J{}_K,
\end{equation}
where $B$ labels the hinge, and the result is independent of the choice of the origin point.

Using Theorem \ref{theom:Matrix-log ln(SAS^-1)} in Appendix \ref{sec:Martix Log}, we can verify that the action \eqref{eq:3D-action 3D simp action on hinge gauge} is gauge invariant and hence can be re-written as
\begin{equation}\label{eq:3D-action 3D simp action on hinge}
\begin{aligned}
^3S_B =&\frac{-1}{2m_B} \! \sum_{i=1}^{m_B}\! \epsilon_{IJ}{}^K(\Lambda(i) l_B(i))^I[{\rm Ln}(\Lambda(i)h_B^{(i)}\Lambda^{-1}(i))]^J{\!}_K\\
=&\frac{-1}{2m_B} \! \sum_{i=1}^{m_B}\! \epsilon_{IJ}{}^K(\Lambda(i) l_B(i))^I[\Lambda(i)({\rm Ln}h_B^{(i)})\Lambda^{-1}(i)]^J{\!}_K\\
=&-\frac{1}{2m_B}\sum_{i=1}^{m_B}\epsilon_{IJ}{}^Kl_B^I(i)({\rm Ln}h_B^{(i)})^J{}_K.
\end{aligned}
\end{equation}
The total action is then given as the sum over all the hinges:
\begin{equation}\label{eq:3D-action 3D simp action A}
^3S[l,h]=-\sum_B\frac{1}{2m_B}\sum_{i=1}^{m_B}\epsilon_{IJ}{}^Kl_B^I(i)({\rm Ln}h_B^{(i)})^J{}_K.
\end{equation}
Thus, the Einstein--Cartan action has been defined on a 3-dimensional simplicial manifold, utilizing edge vectors and the holonomies of curves intersecting the interface as fundamental variables. It should be noted that the expression \eqref{eq:3D-action 3D simp action A} is based on the approximation \eqref{eq:3D-action intR=lnh}. We will show in the next subsection that the same expression for the Einstein--Cartan action can also be obtained by considering the direct relation between the holonomy and the integration of the twisted curvature, given by Eq.~\eqref{eq:non-Abe int scrR-h}.

\subsection{Origin points average}\label{ssec:3D-action B}

Consider again a small region in the 3-dimensional manifold. By choosing a base point $O$ in the region, the twisted curvature 2-form at an arbitrary point $p$ can be defined as
\begin{equation}
\mathscr{R}^I{}_J(p)=(h^{-1})^I{}_K(p)R^K{}_L(p)h^L{}_J(p),
\end{equation}
 and Eq.~\eqref{eq:covariant e} gives the covariant triad as
\begin{equation}
\mathscr{E}^I(p)=(h^{-1})^I{}_J(p)e^J(p),
\end{equation}
where $h^I{}_J(p)$ represents the holonomy along a curve from $O$ to $p$. The Einstein--Cartan action in this region can then be rewritten as
\begin{equation}
\begin{aligned}
^3S[e,\omega]=&\int \frac{1}{2}\epsilon_{IJ}{}^K e^I\wedge R^J{}_K\\
=&\int \frac{1}{2}\epsilon_{IJ}{}^K (h\mathscr{E})^I\wedge (h\mathscr{R}h^{-1})^J{}_K\\
=&\int \frac{1}{2}\epsilon_{IJ}{}^K \mathscr{E}^I\wedge \mathscr{R}^J{}_K,
\end{aligned}
\end{equation}
where we used the fact that the Levi-Civita symbol is invariant under the action of $h$.

Given an internal vector $b^I$ at the base point $O$, one can assign the related internal vector for any other point $p$ by
\begin{equation}\label{eq:3D-action b-hb}
b^I(p)=h^I{}_J(p)b^J(O),
\end{equation}
and hence the corresponding vector field
\begin{equation}
b^a(p)=e^a_I(p)b^I(p).
\end{equation}
Again, the spin connection in the region is assumed to satisfy the condition \eqref{eq:3D-action omega req A}. In the local right-handed coordinate system $\{x,y,z\}$ such that $(\partial/\partial z)^a=b^a$, Eqs. \eqref{eq:3D-action omega-xy} and \eqref{eq:3D-action R-xy} are still valid. Hence the holonomy can be written as
\begin{equation}
h^I{}_J(p)=\mathcal{P}\exp(-\int \omega_a{}^I{}_J(x,y) {\rm d}x^a)=h^I{}_J(x,y),
\end{equation}
and thus the twisted curvature reads
\begin{equation}
\mathscr{R}^I{}_J=\mathscr{R}_{xy}{}^I{}_J(x,y) {\rm d}x \wedge {\rm d}y.
\end{equation}
Moreover, the components of the $b$ field with respect to the covariant triad are constant as
\begin{equation}
\begin{aligned}
\mathscr{E}^I_a(p)b^a(p)=& (h^{-1})^I{}_J(p)e^J_a(p)b^a(p)\\
=& (h^{-1})^I{}_J(p)b^J(p)=b^I(O).
\end{aligned}
\end{equation}
Hence, the Einstein--Cartan action in this region can be decomposed into
\begin{equation}
^3S=\frac{1}{2} \epsilon_{IJ}{}^K \int_l\mathscr{E}_z^I {\rm d}z \iint_S\mathscr{R}_{xy}{}^J{}_K {\rm d}x {\rm d}y,
\end{equation}
where
\begin{equation}
\int_l\mathscr{E}_z^I {\rm d}z =b^I(O)\int_l {\rm d}z=l^I(O).
\end{equation}
As given by Eq.~\eqref{eq:non-Abe int scrR-h}, the non-Abelian Stokes theorem implies
\begin{equation}
\iint_S\mathscr{R}_{xy}{}^J{}_K {\rm d}x{\rm d}y\approx -({\rm Ln}h_C^{(O)})^J{}_K,
\end{equation}
where $h_C^{(O)}$ is the holonomy along a curve surrounding the 2-dimensional integral region $S$ with boundary $C=\partial S$, starting at the point $O$ on $\partial S$. Thus the action can be rewritten as
\begin{equation}\label{eq:3D-action 3S dif to holonomy B}
^3S\approx -\frac{1}{2} \epsilon_{IJ}{}^K l^I(O)({\rm Ln}h_C^{(O)})^J{}_K.
\end{equation}

Now we consider the region containing a hinge in the 3-dimensional simplicial manifold with torsion. Due to the existence of torsion at the interfaces, one can not define a consistent vector field $b^a$ satisfying Eq.~\eqref{eq:3D-action b-hb} in all the simplices. The best we can do is to choose one simplex, say the $i$-th simplex, and ask $b^a$ to be parallel to the hinge inside this simplex. Then the integral of the twisted curvature can be calculated by the holonomy surrounding the hinge, starting from the point inside the $i$-th simplex. This gives a particular expression of the action by employing Eq.~\eqref{eq:3D-action 3S dif to holonomy B} as
\begin{equation}
^3S_B^{(i)}=-\frac{1}{2}\epsilon_{IJ}{}^K l^I(i)({\rm Ln}h_B^{(i)})^J{}_K,
\end{equation}
where $l^I(i)$ is the edge vector of the hinge inside the $i$-th simplex, and $h_B^{(i)}$ is defined as
\begin{equation}
h_B^{(i)}\equiv h_{i-1i}h_{i-2i-1}\cdots h_{12}h_{m1}\cdots h_{i+1i+2}h_{ii+1}.
\end{equation}
The Einstein--Cartan action can be approximated by averaging over all starting points within each simplex as
\begin{equation}
^3S_B=-\frac{1}{2m_B}\sum_{i=1}^{m_B}\epsilon_{IJ}{}^Kl_B^I(i)({\rm Ln}h_B^{(i)})^J{}_K,
\end{equation}
which takes the same form as Eq.~\eqref{eq:3D-action 3D simp action on hinge}. The total action can then be obtained by summing over all the hinges and hence takes the same form as Eq.~\eqref{eq:3D-action 3D simp action A}.

\hspace*{\fill}

To summarize, the same expression \eqref{eq:3D-action 3D simp action A} for the Einstein--Cartan action on the simplicial manifold with torsion has been obtained by two approaches. In the first approach the holonomy can be regarded as coming from the spin connection with the field $b^a$ satisfying the condition \eqref{eq:3D-action omega req A}, and the action corresponds to the classical action integral on a small region in the continuum manifold. However, as shown by Eq.~\eqref{eq:non-Abe int scrR-R}, the approximation \eqref{eq:3D-action intR=lnh} of the curvature integral is not a direct application of the non-Abelian Stokes theorem. In contrast, in the second approach the non-Abelian Stokes theorem of approximating the twisted curvature through holonomy is directly applied, though the discrete action can not be regarded as an integral with respect to a consistent field $b^a$ in a continuous manifold.

In comparison with the first-order formalism of the Regge action given in Ref.~\cite{Caselle:1989hd}, the term ${\rm Ln}h$ is employed in our expression \eqref{eq:3D-action 3D simp action A} instead of $h$. We will show in the following sections that the action \eqref{eq:3D-action 3D simp action A} can be reduced to the Regge action in the torsion-free case and the torsion-free holonomy is the solution to the equation of motion obtained by varying the holonomy. This is not the case for the action expression in Ref.~\cite{Caselle:1989hd}.

\section{Reduction to Regge action in torsion-free case}\label{sec:reduce to tor-free}

In this section, we are going to show that, in the absence of torsion, the Einstein--Cartan action \eqref{eq:3D-action 3D simp action A} in the simplicial manifold can be reduced to the action of Regge calculus \cite{Regge:1961px}.

\begin{figure}[h]
\centering
\includegraphics[width=0.46\textwidth]{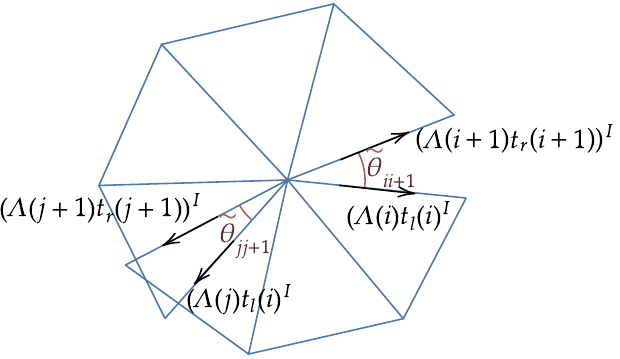}
\caption{2-dimensional projection figure of the tetrahedrons around one hinge map to internal space with triad}
\label{fig:2D proj in internal space}
\end{figure}

In the torsion-free case, the holonomy $\mathring{h}_{ii+1}$ satisfies Eq.~\eqref{eq:torsion-free e-h}, which implies the following equations,
\begin{subequations}\label{eq:tor-free tor-free con to holonomy}
\begin{align}
\label{eq:torsion-free b=hb}b^ae^I_a(i+1)=&\mathring{h}_{ii+1}{}^I{}_Jb^ae^J_a(i),\\
\label{eq:torsion-free t=ht}t^ae^I_a(i+1)=&\mathring{h}_{ii+1}{}^I{}_Jt^ae^J_a(i),\\
\label{eq:torsion-free n=hn}n_ae_I^a(i+1)=&-(\mathring{h}_{ii+1}^{-1}){}^J{}_In_ae_J^a(i),
\end{align}
\end{subequations}
where $b^a$ is the vector parallel to the hinge direction, $t^a$ is a vector parallel to the interface between the $i$-th and the $i+1$-th simplices but perpendicular to the hinge, and $n_a$ is the normal co-vector of the interface. Before proceeding with the computations, we introduce some notation and definitions concerning the wedge product of vectors. Let $u^I$ and $v^I$ be two 3-dimensional internal vectors. We define
\begin{equation}
(u\wedge v)^I{}_J:=u^I v_J-v^I u_J,
\end{equation}
and its norm is given by
\begin{equation}
\|u\wedge v\|=\|{}^*(u\wedge v)\|=\| \epsilon_{IJ}{}^K u^I v^J\|=\|u\|\|v\| \sin\lambda
\end{equation}
where
\begin{equation}
\lambda=\arccos \langle u,v\rangle.
\end{equation}
Consider a region containing the hinge together with all the simplices incident on it in the 3-dimensional simplicial manifold. By choosing an edge vector $\tilde{l}^I$ in certain simplex  as the fiducial one on the hinge, all the other edge vectors in different simplices  can be rotated to $\tilde{l}^I$ by the following gauge transformations:
\begin{equation}
\tilde{e}^I_a(i)=\Lambda^I{}_J(i)e^J_a(i),
\end{equation}
where the group elements of the transformations are given by
\begin{equation}
\Lambda(i)=\exp\left[-\lambda_i \frac{l(i)\wedge \tilde{l}}{\|l(i)\wedge \tilde{l}\|}\right],
\end{equation}
with
\begin{equation}
\lambda_i=\arccos\left\langle l(i),\tilde{l} \right\rangle.
\end{equation}
Here $i$ is used to label the simplices and $l^I(i)$ is the edge vector of the hinge inside the $i$-th simplex. 
After the gauge transformations, all the edge vectors of the hinge are represented by the same fiducial vector $\tilde{l}^I$. Since the torsion-free holonomy maps the hinge vector in one simplex to that in the adjacent simplex by Eq.~\eqref{eq:torsion-free b=hb}, it leaves the common vector $\tilde{l}^I$ invariant. Therefore $\mathring{\tilde{h}}_{ii+1}$ belongs to the $\mathrm{SO}(2)$ subgroup of $\mathrm{SO}(3)$ stabilizing $\tilde{l}^I$, and can be written as the following rotation:
\begin{equation}
\mathring{\tilde{h}}_{ii+1}{}^I{}_J=\exp\left[-\tilde{\theta}_{ii+1}\epsilon^I{}_{JK}\hat{l}^K\right],
\end{equation}
where $\hat{l}^I:=\frac{\tilde{l}^I}{\|\tilde{l}\|}$ is the unit vector along the hinge. To specify its rotation angle $\tilde{\theta}_{ii+1}$ satisfying Eqs. \eqref{eq:torsion-free t=ht} and \eqref{eq:torsion-free n=hn}, let $t_l^I(i)$ and $t_r^I(i+1)$ be the unit internal vectors tangent to the common interface between the $i$-th and the $(i+1)$-th simplices and perpendicular to the hinge, as viewed from the two simplices, respectively. The subscripts $l$ and $r$ distinguish the two tangent directions associated with the two faces of one simplex. The rotation angle $\tilde{\theta}_{ii+1}$is then the oriented angle between these two vectors in the plane perpendicular to $\tilde{l}^I$ given by
\begin{equation}
\begin{aligned}
\tilde{\theta}_{ii+1}\!=&(-1)^s\arccos\left\langle \Lambda(i)t_l(i),\Lambda(i+1)t_r(i+1) \right\rangle\\
=&(-1)^s\arccos\bigl(t_{rK}(i\!+\!1)(\Lambda^{-1})^K{\!}_I(i\!+\!1)\Lambda^I{\!}_J(i)t_l^J(i)\bigr),
\end{aligned}
\end{equation}
where $s$ fixes the orientation of the rotation with respect to the hinge direction as
\begin{equation}
s\!=\!\begin{cases}
0,&\!\!\!\!\!\!\! \text{ if } \epsilon_{IJK}(\Lambda(i)t_l(i))^I \!\wedge\! (\Lambda(i+1)t_r(i+1))^J\tilde{l}^K \!>\! 0 \\
1,&\!\!\!\!\!\!\! \text{ if } \epsilon_{IJK}(\Lambda(i)t_l(i))^I \!\wedge\! (\Lambda(i+1)t_r(i+1))^J\tilde{l}^K \!<\! 0
\end{cases}\!.
\end{equation}
For instance, as illustrated in Fig.~\ref{fig:2D proj in internal space}, $\tilde{\theta}_{ii+1}$ is positive ($s=0$), while $\tilde{\theta}_{jj+1}$ is negative ($s=1$).
The torsion-free holonomy of spin connection before the gauge transformation can then be written as
\begin{equation}
\begin{aligned}
\mathring{h}_{ii+1}=&\Lambda^{-1}(i+1)\mathring{\tilde{h}}_{ii+1}\Lambda(i)\\
=&\exp\bigl[\lambda_{i+1} \frac{l(i+1)\wedge \tilde{l}}{\|l(i+1)\wedge \tilde{l}\|}\bigr]\\
&\times \exp\bigl[-\tilde{\theta}_{ii+1}\epsilon^I{}_{JK}\hat{l}^K\bigr]\exp\bigl[-\lambda_i \frac{l(i)\wedge \tilde{l}}{\|l(i)\wedge \tilde{l}\|}\bigr].
\end{aligned}
\end{equation}
The holonomy along a closed curve surrounding the hinge, starting from a point inside the $i$-th simplex, reads
\begin{equation}\label{eq:tor-free tor-free holonomy}
\begin{aligned}
\mathring{h}_B^{(i)}=&\Lambda^{-1}(i)\exp\left[-\tilde{\theta}_{i-1i}U\right]\exp\left[-\tilde{\theta}_{i-2i-1}U\right]\cdots\\
&\times\!\exp\left[-\tilde{\theta}_{12}U\right]\exp\left[-\tilde{\theta}_{m1}U\right]\!\cdots\exp\left[-\tilde{\theta}_{ii+1}U\right]\Lambda(i)\\
=&\Lambda^{-1}(i)\exp\left[-\sum_{i=1}^m\tilde{\theta}_{ii+1}U\right]\Lambda(i)\\
=&\Lambda^{-1}(i)\exp\left(-\varepsilon U\right)\Lambda(i),
\end{aligned}
\end{equation}
where we denote
\begin{equation}
U^I{}_J=\epsilon^I{}_{JK}\hat{l}^K,
\end{equation}
and the deficit angle reads $\varepsilon=\sum_{i=1}^m\tilde{\theta}_{ii+1}$. Substituting the torsion-free holonomy \eqref{eq:tor-free tor-free holonomy} into the action \eqref{eq:3D-action 3D simp action on hinge}, we obtain
\begin{equation}\label{eq:tor-free action-B}
\begin{aligned}
^3\!S_B\!=&\frac{-1}{2m_B}\!\sum_{i=1}^{m_B}\!\epsilon_{IJ}{}^Kl_B^I(i)[\Lambda^{-1}(i){\rm Ln}(\exp\left(-\varepsilon U\right))\Lambda(i)]^J{\!}_K\\
=&-\frac{1}{2}\epsilon_{IJ}{}^Kl_B^I(i)[\Lambda^{-1}(i)\left(-\varepsilon U\right)\Lambda(i)]^J{}_K\\
=&\frac{1}{2}\epsilon_{IJ}{}^K\tilde{l}_B^I\varepsilon \epsilon^J{}_{KL}\hat{l}_B^L\\
=&\varepsilon L,
\end{aligned}
\end{equation}
where $L=\|l\|$ is the length of the edge (i.e., the hinge in 3-dimensions). Summing over all the hinges with Eq.~\eqref{eq:tor-free action-B} gives the Regge action.

\section{Extension to 4-dimensional Lorentzian simplicial manifold}\label{sec:4D-expand}

In this section, we extend the Einstein--Cartan action \eqref{eq:3D-action 3D simp action A} to 4-dimensional Lorentzian simplicial manifold. Recall that the Einstein--Cartan action on a 4-dimensional differential manifold is given by
\begin{subequations}
\begin{gather}
^4S_{EC}[e,\omega]=\int\frac{1}{4}\epsilon_{IJK}{}^Le^I\wedge e^J\wedge R^K{}_L,\\
\shortintertext{or equivalently,}
^4S_{EC}[e,\omega]=\int\frac{1}{4}\epsilon_{IJK}{}^L\mathscr{E}^I\wedge \mathscr{E}^J\wedge \mathscr{R}^K{}_L.
\end{gather}
\end{subequations}
In the 4-dimensional simplicial manifold, the hinge is a 2-dimensional simplex, i.e. a triangle. Additionally, the non-vanishing curvature that contributes to the action should be condensed on the hinges. Similarly to the 3-dimensional simplicial manifold, the action in the 4-dimensional simplicial manifold can be obtained as
\begin{equation}\label{eq:4D-expand 4D simp action}
^4S[l,h]=\sum_B\frac{1}{2m_B}\sum_{i=1}^{m_B}{\rm tr}(A_B(i){\rm Ln}h_B^{(i)}),
\end{equation}
with
\begin{equation}\label{eq:4D-expand A(i)}
\begin{aligned}
(A_B)_{KL}(i)=& \frac{1}{2}\epsilon_{IJKL}l_{B1}^I(i)l_{B2}^J(i)\\
=& \frac{1}{2}\epsilon_{IJKL}l_{B2}^I(i)l_{B3}^J(i)\\
=& \frac{1}{2}\epsilon_{IJKL}l_{B3}^I(i)l_{B1}^J(i),
\end{aligned}
\end{equation}
where $l_{Bp}^I(i),\ p=1,2,3$ represent the three edge vectors of the 2-dimensional hinge in the $i$-th simplex, satisfying
\begin{equation}\label{eq:4D-expand clo-condition dir}
\sum_{p=1}^3l_{Bp}^I(i)=0,
\end{equation}
and the holonomy is given by
\begin{equation}
h_B^{(i)}\equiv h_{i-1i}h_{i-2i-1}\cdots h_{12}h_{m1}\cdots h_{i+1i+2}h_{ii+1}.
\end{equation}
Here the simplices surrounding the hinge $B$ are labeled in the sequence $i=1,2,\dots,m_B$ such that their ordering, together with the bivector $A_B(i)$, defines a right-handed orientation, and the loop defining $h_B^{(i)}$ follows the same orientation.

It should be noted that here the holonomy $h_{ii+1}$ belongs to the proper Lorentz group $\mathrm{SO}^+(1,3)$. $\mathrm{SO}^+(1,3)$ group is not compact; the exponential map from the corresponding Lie algebra to the Lie group is not an onto map. As a result, the logarithm of a group element does not necessarily lie in its Lie algebra. However, the logarithm of the group element, when considered in the matrix representation, can still be defined as the inverse of the exponential map. Furthermore, the matrix representation of the group $\mathrm{SO}^+(1,3)$ has no negative or zero real eigenvalues. This property ensures the existence of the real principal branch of the logarithm. The detailed discussion on the logarithm in the matrix representation is given in Appendix \ref{sec:Martix Log}.

It should also be noted that the gauge transformations of the theory are restricted to the $\mathrm{SO}^+(1,3)$ group. This implies that the edge vectors, which point in the same direction from the starting vertex to the target vertex in a 3-simplex belonging to two 4-simplices, have to be identical up to an $\mathrm{SO}^+(1,3)$ gauge transformation. Consequently, for an edge belonging to different 4-simplices, its timelike edge vectors in the different simplices have to be consistent in time direction. This means that these timelike vectors should be either all future-pointed or all past-pointed. Additionally, it is required that the edge vectors be non-null.

\section{Equations of motion on the Simplicial Manifold}\label{sec:EoM}

In this section, we derive the equations of motion by the variation of the action \eqref{eq:4D-expand 4D simp action} and explore their implications. In the first part of this section, we calculate the variation of the action with respect to the edge vectors $l_p^I$, and show that the resulting equation corresponds to the equation of motion derived from the variation of the tetrad in the continuum theory. In the second part, we calculate the variation of the action with respect to the holonomies $h_e{}^I{}_J$ passing across the interfaces, and show that the torsion-free holonomy is a solution of the resulting equation of motion.

\subsection{Variation of edge vectors}\label{ssec:EoM-edge vector}

\begin{figure}[h]
\centering
\includegraphics[width=0.3\textwidth]{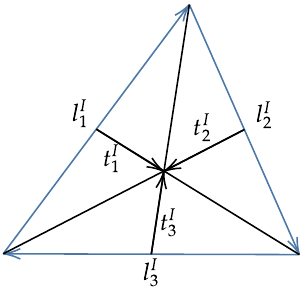}
\caption{Hinge(triangle) with centroid}
\label{fig:hinge with centroid}
\end{figure}

Under the closure condition \eqref{eq:4D-expand clo-condition dir}, $A_B$ can be equivalently expressed in three forms given in Eq.~\eqref{eq:4D-expand A(i)}. To render the equations of motion symmetric with respect to different edge vectors, we express $A_B$ as the average of the three forms. Thus, the variation of $A_B(i)$ with respect to the edge vectors reads
\begin{equation}
\begin{aligned}
\delta (A_B)_{KL}(i) \!=&\frac{1}{6}\epsilon_{IJKL}\bigl((\delta l_1^I(i))l_2^J(i)+l_1^I(i)\delta l_2^J(i)\\
& +(\delta l_2^I(i))l_3^J(i)+l_2^I(i)\delta l_3^J(i)\\
& +(\delta l_3^I(i))l_1^J(i)+l_3^I(i)\delta l_1^J(i)\bigr)\\
=&\frac{1}{6}\epsilon_{IJKL}\bigl[(\delta l_1^I(i))(l_2^J(i)-l_3^J(i))\\
& +(\delta l_2^I(i))(l_3^J(i)-l_1^J(i))\\
& +(\delta l_3^I(i))(l_1^J(i)-l_2^J(i))\bigr]\\
=& \epsilon_{IJKL}[t_1^J(i)\delta l_1^I(i) \!+\! t_2^J(i)\delta l_2^I(i) \!+\! t_3^J(i)\delta l_3^I(i)]\\
=& \epsilon_{IJKL}\sum_{p=1}^3t_p^J(i)\delta l_p^I(i)
\end{aligned}
\end{equation}
where $t_p^I$ is the vector pointing from the midpoint of edge $l_p$ to the centroid of the triangle, as shown in Fig.~\ref{fig:hinge with centroid}.
Then the variation of the action \eqref{eq:4D-expand 4D simp action} with respect to the edge vectors reads as
\begin{equation}
\begin{aligned}
\delta (^4S)(l_p^I)=&\sum_B \frac{1}{2m_B}\sum_{i=1}^{m_B}{\rm tr}[(\delta A_B(i)){\rm Ln}h_B^{(i)}]\\
=&\sum_B \frac{1}{2m_B}\sum_{i=1}^{m_B}\epsilon_{IJK}{}^L\sum_{p=1}^3t_{Bp}^J(i)\\
& \times ({\rm Ln}h_{B,\epsilon}^{(i)})^K{}_L\delta l_{Bp}^I(i).
\end{aligned}
\end{equation}
Here, the sign $\epsilon=\pm1$ specifies the orientation of the holonomy associated with each pair $(B,p)$. We take $\epsilon=+1$ when the orientation encoded by $l_{Bp}(i)\wedge t_{Bp}(i)$ agrees with that of $A_B(i)$, and $\epsilon=-1$ when it is opposite. Accordingly, we define
\begin{equation}\label{eq:EoM h-epsilon}
h_{B,\epsilon}^{(i)}:=\left\{\begin{matrix}
h_B^{(i)}, & \epsilon=+1,\\
(h_B^{(i)})^{-1}, & \epsilon=-1,
\end{matrix}\right.
\end{equation}
such that $l_{Bp}^I(i)$, $t_{Bp}^J(i)$, and the oriented loop defining $h_{B,\epsilon}^{(i)}$ follow the right-hand convention.
In the region containing the simplices sharing the common edge $l_p$, one can always find suitable gauge transformations to rotate the edge vectors of different simplices to a chosen fiducial one as
\begin{equation}\label{eq:EoM lp-gauge}
\Lambda_{Bi}{}^I{}_Jl_{Bp}^J(i)=\tilde{l}_p^I.
\end{equation}
By assuming that the variation commutes with gauge transformations, we obtain
\begin{equation}
\begin{aligned}
\delta (^4S)(l_p^I)=&\sum_B \frac{1}{2m_B}\sum_{i=1}^{m_B}\epsilon_{IJK}{}^L\sum_{p=1}^3t_{Bp}^J(i)\\
& \times ({\rm Ln}h_{B,\epsilon}^{(i)})^K{}_L (\Lambda_{Bi}^{-1})^I{}_M\delta \tilde{l}_{p}^M\\
=&\sum_p\sum_B \frac{1}{2m_B}\sum_{i=1}^{m_B}\epsilon_{IJK}{}^Lt_{Bp}^J(i)\\
& \times ({\rm Ln}h_{B,\epsilon}^{(i)})^K{}_L (\Lambda_{Bi}^{-1})^I{}_M\delta \tilde{l}_{p}^M.
\end{aligned}
\end{equation}
So the corresponding equation of motion is given by
\begin{subequations}\label{eq:EoM EoM-edge}
\begin{gather}
\label{eq:EoM EoM-edge gau-tran}
\sum_{B\in \partial^*p} \frac{1}{2m_B}\sum_{i=1}^{m_B}\epsilon_{IJK}{}^Lt_{Bp}^J(i)({\rm Ln}h_{B,\epsilon}^{(i)})^K{}_L(\Lambda_{Bi}^{-1})^I{}_M=0,\\
\shortintertext{i.e.,}
\label{eq:EoM EoM-edge gau-cho}
\sum_{B\in \partial^*p}\! \frac{1}{2m_B}\sum_{i=1}^{m_B}\epsilon_{IJK}{}^L\tilde{t}_{Bp}^J(i)({\rm Ln}\tilde{h}_{B,\epsilon}^{(i)})^K{}_L=0,
\end{gather}
\end{subequations}
where $\partial^*p$ represents the region of simplices sharing the common edge $l_p$, $B\in \partial^*p$ labels the hinges sharing the common edge $l_p$, $\tilde{t}_{Bp}^I(i)$ and $\tilde{h}_B^{(i)}{}^K{}_L$ are given by
\begin{equation}\label{eq:EoM t-gauge}
\tilde{t}_{Bp}^I(i)=\Lambda_{Bi}{}^I{}_Jt_{Bp}^J(i),
\end{equation}
and
\begin{equation}\label{eq:EoM h-gauge}
\tilde{h}_B^{(i)}{}^K{}_L=\Lambda_{Bi}{}^K{}_Mh_B^{(i)}{}^M{}_N(\Lambda_{Bi}^{-1})^N{}_L,
\end{equation}
together with the closure condition \eqref{eq:clo-condition l} of the triangles.

Recall that in the continuum theory, the variation of the Einstein--Cartan action with respect to the tetrad gives the following equation of motion:
\begin{equation}\label{eq:EoM Einstein continuous}
E^I{}_a\equiv R^{b}{}_{a}e^I_b-\frac{1}{2}Re^I_a=0.
\end{equation}
where $R_{ab}=e^c_I e_b^J R_{ca}{}^I{}_J$ and $R=e^a_I e^{bI} R_{ab}$. Note that Eq.~\eqref{eq:EoM Einstein continuous} is not yet the vacuum Einstein's equation since the two indices of  $E_{ab}$ is not symmetric. For later convenience, we introduce the vector-valued 3-form \cite{Barrett:1985nc}:
\begin{equation}\label{eq:EoM G 1-form}
G^I:=\frac{1}{2} e^J\wedge R^K{}_L\epsilon_{JK}{}^{LI},
\end{equation}
such that
\begin{equation}
E^I{}_d=\frac{1}{3!} G^{Iabc}\epsilon_{abcd}.
\end{equation}

To compare the discrete equations with Eq.~\eqref{eq:EoM Einstein continuous}, we need a suitable integral of the corresponding 3-form \eqref{eq:EoM G 1-form}.
In a 4-dimensional differential manifold, the covariant integration of Eq.~\eqref{eq:EoM G 1-form} over a 3-dimensional region $V$ can be defined as
\begin{equation}\label{eq:EoM int G con}
\begin{aligned}
\mathcal{G}^I=&\int_V (h(O,x)^{-1})^I{}_JG^J(x)\\
=&\int_V \frac{1}{2}(h(O,x)^{-1})^I{}_Me^J(x)\wedge R^K{}_L(x)\epsilon_{JK}{}^{LM}\\
=&\int_V \frac{1}{2}(h(O,x)^{-1})^I{}_M(h(O,x)\mathscr{E}(x))^J\\
& \wedge (h(O,x)\mathscr{R}(x)h(O,x)^{-1})^K{}_L\epsilon_{JK}{}^{LM}\\
=&\int_V \frac{1}{2}\mathscr{E}^J(x)\wedge \mathscr{R}^K{}_L(x) \epsilon_{JK}{}^{LI},
\end{aligned}
\end{equation}
where $h(O,x)$ represents the holonomy along the curve from the starting point $O$ to the target point $x$. Obviously, this integral depends on the choice of the starting point $O$ and is covariantly transformed at $O$ under a gauge transformation.

We now evaluate the contribution to Eq.~\eqref{eq:EoM int G con} from a single hinge in the 4-dimensional simplicial manifold. Consider a 3-dimensional region $V$ intersecting one hinge $B$ at a curve $s$. After choosing the starting point inside the $i$-th simplex, similarly to the treatment in Sec.~\ref{ssec:3D-action B}, the integral $\mathcal{G}^I(i)$ corresponding to Eq.~\eqref{eq:EoM int G con} can be approximated as
\begin{equation}\label{eq:EoM int-G(B)}
\begin{aligned}
\mathcal{G}_{(B)}^I(i)\approx& \frac{1}{2}\int_s\mathscr{E}_a^J {\rm d}x^a \iint_S \mathscr{R}_{bc}{}^K{}_L\epsilon_{JK}{}^{LI} {\rm d}x^b {\rm d}x^c\\
\approx& -\frac{1}{2}s^J(i)({\rm Ln}h_B^{(i)})^K{}_L\epsilon_{JK}{}^{LI},
\end{aligned}
\end{equation}
where $s^J(i)$ is the internal displacement vector pointing from the origin to the target point of the curve $s$ within the $i$-th simplex.

\begin{figure}[h]
\centering
\includegraphics[width=0.3\textwidth]{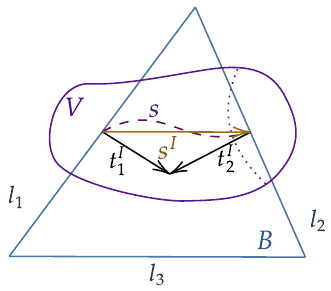}
\caption{Hinge B intersected by a 3-dimensional region V}
\label{fig:hinge B with V}
\end{figure}

To make the correspondence between Eq.~\eqref{eq:EoM int-G(B)} and the continuum equation more explicit, consider a 3-dimensional region $W$ whose characteristic size is much larger than the simplicial scale, so that it intersects a large number of edges, while remaining sufficiently small on the macroscopic scale. We choose $W$ such that its intersections with these edges occur at their midpoints. 
For each hinge $B$ intersected by $W$, the portion of $W$ in a neighborhood of $B$ plays the role of the region $V$ considered above. In particular, suppose that $W\cap B$ is a curve $s$ whose endpoints lie at the midpoints of two boundary edges $l_1$ and $l_2$ of $B$, as shown in Fig.~\ref{fig:hinge B with V}. Then Eq.~\eqref{eq:EoM int-G(B)} can be employed to approximate the continuous integration \eqref{eq:EoM int G con} in this region. Additionally, the vector $s^J(i)$ can be decomposed as:
\begin{equation}\label{eq:EoM s=t-t}
s^J(i)=t_1^J(i)-t_2^J(i),
\end{equation}
where $t_p^I$ is the vector pointing from the midpoint of edge $l_p$ to the centroid of the triangle as shown in Fig.\ref{fig:hinge B with V}. Here we use the centroid of the triangular hinge to define this decomposition. This differs from the construction in Ref.~\cite{Barrett:1985nc}, where the corresponding decomposition in the torsion-free case is based on the circumcenter of the hinge. Then Eq.~\eqref{eq:EoM int-G(B)} can be decomposed into the following two parts:
\begin{equation}
\mathcal{G}^I_{(B)}(i)=\mathcal{G}_{B1}^I(i)+\mathcal{G}_{B2}^I(i).
\end{equation}
For the first term, we retain the orientation of the holonomy $h_B^{(i)}$ used in Eq.~\eqref{eq:EoM int-G(B)}, so that
\begin{equation}
\mathcal{G}_{B1}^I(i)=-\frac{1}{2}t_1^J(i)({\rm Ln}h_B^{(i)})^K{}_L\epsilon_{JK}{}^{LI}.
\end{equation}
For the second term, the relative minus sign arising from Eq.~\eqref{eq:EoM s=t-t} can instead be absorbed by reversing the orientation of the holonomy around the hinge. Indeed, reversing the orientation gives ${\rm Ln}((h_B^{(i)})^{-1})=-{\rm Ln}h_B^{(i)}$ and hence
\begin{equation}
\mathcal{G}_{B2}^I(i)=-\frac{1}{2}t_2^J(i)({\rm Ln}((h_B^{(i)})^{-1}))^K{}_L\epsilon_{JK}{}^{LI}.
\end{equation}
For each edge $l_p$ intersected by $W$, we first orient $l_p$ consistently toward the same side of $W$. The orientation of the holonomy around the hinge $B$ is then fixed by requiring it to satisfy the right-hand rule with $t_{Bp}^J(i)$. Using the orientation convention introduced above in Eq.~\eqref{eq:EoM h-epsilon}, the two contributions can be written uniformly as
\begin{equation}\label{eq:EoM GBp(i)}
\mathcal{G}_{Bp}^I(i)=-\frac{1}{2}t_{Bp}^J(i)({\rm Ln}h_{B,\epsilon}^{(i)})^K{}_L\epsilon_{JK}{}^{LI},\ p=1,2.
\end{equation}
In this way, the integrated contribution of each hinge is decomposed into the terms naturally associated with the two boundary edges on which the endpoints of $W\cap B$ lie.

The discrete equation \eqref{eq:EoM EoM-edge} obtained by varying with respect to the edge vector, however, involves all hinges incident on that edge rather than an individual hinge. The next step is to combine the contributions associated with the same edge $p$ arising from different hinges. To compare the covariant integral of the 3-form $G^I$ with the discrete equation \eqref{eq:EoM EoM-edge} associated with the edge $p$, Eq.~\eqref{eq:EoM GBp(i)} has to be expressed in the same gauge as that used in Eq.~\eqref{eq:EoM lp-gauge}. Besides, there is still an arbitrary choice of the simplex in which the starting point for Eq.~\eqref{eq:EoM GBp(i)} is taken. Thus, we average the contribution $\mathcal{G}_{Bp}^I(i)$ over the different choices of the starting point, taken inside each of the $m_B$ simplices surrounding the hinge $B$. 
In the gauge of Eq.~\eqref{eq:EoM lp-gauge}, this averaged contribution reads
\begin{equation}\label{eq:EoM GBp}
\begin{aligned}
\mathscr{G}_{Bp}^I=& \frac{1}{m_B}\sum_{i=1}^{m_B}\tilde{\mathcal{G}}_{Bp}^I(i)\\
=& -\frac{1}{2m_B}\sum_{i=1}^{m_B}\tilde{t}_{Bp}^J(i)({\rm Ln}\tilde{h}_{B,\epsilon}^{(i)})^K{}_L\epsilon_{JK}{}^{LI} 
\end{aligned}
\end{equation}
where $\tilde{t}_{Bp}^I(i)$ and $\tilde{h}_B^{(i)}{}^I{}_J$ are given as Eqs.~\eqref{eq:EoM t-gauge} and \eqref{eq:EoM h-gauge}.
We then sum over Eq.~\eqref{eq:EoM GBp} for the hinges sharing the common edge $p$ and obtain
\begin{equation}\label{eq:EoM Gp}
\begin{aligned}
\mathscr{G}_{p}^I=& \sum_{B\in\partial^*p}\mathscr{G}_{Bp}^I\\
=& -\sum_{B\in\partial^*p}\frac{1}{2m_B}\sum_{i=1}^{m_B}\tilde{t}_{Bp}^J(i)({\rm Ln}\tilde{h}_{B,\epsilon}^{(i)})^K{}_L\epsilon_{JK}{}^{LI},
\end{aligned}
\end{equation}
which is the same as the left-hand side of the discrete equation \eqref{eq:EoM EoM-edge gau-cho}. The covariant integral of $G^I$ over $W$ can then be approximated by summing the contributions associated with all the edges intersected by $W$ as
\begin{equation}
\mathscr{G}^I(W)\approx \sum_{p:p\cap W\ne \varnothing}\mathscr{G}_p^I,
\end{equation}
up to contributions associated with the boundary of $W$. Hence, the discrete equation \eqref{eq:EoM EoM-edge gau-cho} implies
\begin{equation}\label{eq:EoM G(W)=0}
\mathscr{G}^I(W) =0,
\end{equation}
up to boundary contributions. In this sense, the discrete equation \eqref{eq:EoM EoM-edge} implies the covariant integral of the continuum tetrad equation. It remains to examine the converse implication.

Suppose that the continuum tetrad equation \eqref{eq:EoM Einstein continuous} holds. Then the covariant integral \eqref{eq:EoM int G con} vanishes for any admissible 3-dimensional region. Consider a 3-dimensional region $V$ as above, whose intersection with a triangular hinge $B$ is a curve $s$. Within the simplicial approximation of Eq.~\eqref{eq:EoM int-G(B)}, the corresponding hinge contribution therefore vanishes for any choice of $s^J(i)$ lying in the tangent plane of the hinge $B$ represented in the internal frame of simplex $i$. Since $t^J_p(i)$ also lies in this tangent plane, we have
\begin{equation}\label{eq:EoM GBp(i)=0}
\mathcal{G}_{Bp}^I(i)=0
\end{equation}
for any $p$. Since these quantities are the elementary terms from which the discrete equation \eqref{eq:EoM EoM-edge} is constructed, \eqref{eq:EoM GBp(i)=0} implies that the corresponding gauge-transformed and $i$-averaged terms also vanish, and hence the discrete edge equation is satisfied.

It should be noted that, in the discrete theory, the equation of motion does not vanish for each individual hinge $B$. Instead, it only vanishes the sum of the contributions from all hinges sharing the common edge $p$. Consequently, within the simplicial approximation adopted above, the continuum tetrad equation implies the discrete edge equation. However, at the level of these integrated hinge contributions, the discrete equation imposes a weaker condition than the continuum tetrad equation: it only vanishes the sum of the contributions associated with all the edges in a region in the sense of Eq.~\eqref{eq:EoM G(W)=0}.

\subsection{Variation of holonomies}\label{ssec:EoM-holonomy}

To vary the action with respect to the holonomies across the interfaces, we need to handle the logarithm of the hinge holonomy appearing in the action \eqref{eq:4D-expand 4D simp action}. For this purpose, we need to use the Taylor expansion in Eq.~\eqref{eq:Matrix-log ln(A)-Taylor series}, and therefore assume the triangulation to be sufficiently fine so that the holonomy around each hinge satisfies
\begin{equation}\label{eq:EoM rho(h-1)<1}
\rho(h_B^{(i)}-I)<1,
\end{equation}
where $\rho$ denotes the spectral radius. This condition ensures that the hinge holonomy lies sufficiently close to the identity. In the four-dimensional defining representation, the nontrivial eigenvalues of the hinge holonomy associated with its boost and rotation sectors can be written as $e^{\pm\eta}$ and $e^{\pm i\theta}$, respectively, where $\eta$ is the boost rapidity and $\theta$ is the rotation angle. The above condition therefore requires
\begin{equation}
|e^{\pm\eta}-1|<1,\qquad |e^{\pm i\theta}-1|<1,
\end{equation}
which, with $\theta$ chosen in the principal range $-\pi<\theta\leq\pi$, gives
\begin{equation}
|\eta|<\ln 2,\qquad |\theta|<\frac{\pi}{3}.
\end{equation}
Under this assumption, the Taylor expansion of ${\rm Ln}h_B^{(i)}$ converges and can be used in the variation of the action \eqref{eq:4D-expand 4D simp action}. Moreover, this condition is independent of the gauge choice. Indeed, a gauge transformation in the $i$-th simplex acts on the hinge holonomy as
\begin{equation}
h_B^{(i)}\mapsto \Lambda(i)h_B^{(i)}\Lambda(i)^{-1},
\end{equation}
which leaves its eigenvalues unchanged. So the convergence condition \eqref{eq:EoM rho(h-1)<1} is independent of the gauge choice.

With the convergence condition \eqref{eq:EoM rho(h-1)<1}, Eq.~\eqref{eq:Matrix-log del(lnA)-Taylor series} can be applied to the logarithm appearing in each hinge contribution to the action \eqref{eq:4D-expand 4D simp action}. Keeping $A_B$ fixed, the variation of a single hinge term with respect to $h_B$ reads
\begin{subequations}
\begin{gather}
\begin{aligned}
\delta({\rm tr}(A_B{\rm Ln}h_B))(h_B)=&\frac{1}{2}{\rm tr}\left[(h_B^{-1}A_B+A_Bh_B^{-1})\delta h_B\right]\\
&+\frac{1}{2}{\rm tr}\bigl[\sum_{q=0}^\infty\sum_{p=0}^{\infty}c_{pq}(h_B-I)^p\\
& \times [h_B,[A_B,h_B]](h_B-I)^q\delta h_B\bigr]
\end{aligned}
\shortintertext{where}
c_{pq}=\frac{(-1)^{p+q}(p+1)(q+1)}{p+q+3}.
\end{gather}
\end{subequations}
The variation of the action \eqref{eq:4D-expand 4D simp action} with respect to the interface holonomies can then be written as
\begin{widetext}
\begin{equation}
\begin{aligned}
\delta(^4S)(h_e)=&\sum_B\frac{1}{4m_B}\sum_{i=1}^{m_B}\{{\rm tr}\left[((h_B^{(i)})^{-1}A_B(i)+A_B(i)(h_B^{(i)})^{-1})\delta h_B^{(i)}\right]\\
&+{\rm tr}\left[\sum_{q=0}^\infty\sum_{p=0}^{\infty}c_{pq}(h_B^{(i)}-I)^p[h_B^{(i)},[A_B(i),(h_B^{(i)})]](h_B^{(i)}-I)^q\delta h_B^{(i)}\right]\}\\
=&\sum_B\frac{1}{4m_B}\sum_{i=1}^{m_B}\{\left(\vec{\prod}_{l=j_B-1}^{i_B}h_{ll+1}\right)[(h_B^{(i)})^{-1}A_B(i)+A_B(i)(h_B^{(i)})^{-1}\\
&+\sum_{q=0}^\infty\sum_{p=0}^{\infty}c_{pq}(h_B^{(i)}-I)^p[h_B^{(i)},[A_B(i),(h_B^{(i)})]](h_B^{(i)}-I)^q]\left(\vec\prod_{k=i_B-1}^{j_B+1}h_{kk+1}\right)\}{}^I{}_J\delta h_{j_Bj_B+1}{}^J{}_I\\
=&\sum_e\sum_{B=1}^4 \frac{1}{4m_B}\sum_{i_B=1}^{m_B}\{\left(\vec{\prod}_{l=j_B-1}^{i_B}h_{ll+1}\right)[(h_B^{(i)})^{-1}A_B(i_B)+A_B(i_B)(h_B^{(i)})^{-1}\\
&+\sum_{q=0}^\infty\sum_{p=0}^{\infty}c_{pq}(h_B^{(i)}-\mathbf{1})^p[h_B^{(i)},[A_B(i_B),h_B^{(i)}]](h_B^{(i)}-\mathbf{1})^q]\left(\vec\prod_{k=i_B-1}^{j_B+1}h_{kk+1}\right)\}^I{}_J\delta h_e{}^J{}_I,
\end{aligned}
\end{equation}
\end{widetext}
where $e$ labels a curve crossing a specific interface, $h_e$ denotes the corresponding holonomy identified with $h_{j_Bj_B+1}$, and the symbol $\vec{\prod}$ is defined by
\begin{equation}
\vec\prod_{k=i}^jh_{kk+1}:= h_{ii+1}h_{i-1i}\cdots h_{j+1j+2}h_{jj+1}.
\end{equation}
Therefore, the equation of motion associated with the holonomy $h_e$ takes the form
\begin{widetext}
\begin{equation}\label{eq:EoM EoM-holonomy}
\begin{aligned}
&\sum_{B=1}^4 \frac{1}{4m_B}\sum_{i_B=1}^{m_B}\{\left(\vec{\prod}_{l=j_B-1}^{i_B}h_{ll+1}\right)[(h_B^{(i)})^{-1}A_B(i)+A_B(i)(h_B^{(i)})^{-1}\\
&+\sum_{q=0}^\infty\sum_{p=0}^{\infty}c_{pq}(h_B^{(i)}-\mathbf{1})^p[h_B^{(i)},[A_B(i),h_B^{(i)}]](h_B^{(i)}-\mathbf{1})^q]\left(\vec\prod_{k=i_B-1}^{j_B+1}h_{kk+1}\right)\}^I{}_J=0.
\end{aligned}
\end{equation}
\end{widetext}

Though it is not easy to solve Eq.~\eqref{eq:EoM EoM-holonomy}, we check whether the torsion-free holonomy is a solution of this equation. In the torsion-free situation, $\mathring{h}_B^{(i)}$ should commute with $A_B(i)$ as matrices, i.e.,
\begin{equation}
[A_B(i),\mathring{h}_B^{(i)}]=0.
\end{equation}
Using this relation, for torsion-free holonomies the left-hand side of Eq.~\eqref{eq:EoM EoM-holonomy} reduces to
\begin{widetext}
\begin{equation}\label{eq:EoM EoM-holonomy tor-free reduce}
\sum_{B=1}^4 \frac{1}{2m_B}\sum_{i_B=1}^{m_B}[\left(\vec{\prod}_{l=j_B-1}^{i_B}\mathring{h}_{ll+1}\right)A_B(i)(\mathring{h}_B^{(i)})^{-1}\left(\vec\prod_{k=i_B-1}^{j_B+1}\mathring{h}_{kk+1}\right)]^I{}_J=\frac{1}{2}\sum_{B=1}^4A_B{}^I{}_K(j_B)(\mathring{h}_e^{-1})^K{}_J=0,
\end{equation}
\end{widetext}
where the following closure condition for the bivectors $A_B(j_B)$ on the interface is used in the last step:
\begin{equation}
\sum_{B=1}^4A_B{}^I{}_J(j_B)=0.
\end{equation}
Therefore the torsion-free holonomy is a solution to the discrete equation of motion \eqref{eq:EoM EoM-holonomy}.

\section{Conclusion}\label{sec:conslusion}

In previous sections, the discrete Riemann-Cartan geometry has been constructed on an $n$-dimensional simplicial manifold, on which the corresponding simplicial formulation of Einstein--Cartan gravity can be developed. The discrete geometry is described by edge vectors and holonomies across interfaces, playing the roles of discrete tetrads and spin connections, respectively. Within this framework, curvature is localized on hinges, while torsion is concentrated on interfaces, providing a discrete counterpart of the continuum Riemann-Cartan geometry. The geometric role of torsion in simplicial geometry is clarified, and its relation to the Burgers vector in regions of vanishing Riemann curvature is also established. This framework allows us to formulate a simplicial Einstein--Cartan action and to establish its relation to Regge calculus in the torsion-free sector. 

By employing the non-Abelian Stokes theorem, the simplicial action for Einstein--Cartan gravity has been obtained in three or four dimensions with Euclidean and Lorentzian signatures respectively, as a discrete analogue of the continuous Einstein--Cartan action. It is shown that this action reduces to the Regge action in the torsion-free case, thereby recovering the traditional Regge calculus. Moreover, the equations of motion of the 4-dimensional discrete gravity have been derived by varying the action with respect to the edge vectors and the holonomies, respectively. The consistency between the discrete equation obtained by varying the edge vectors and the corresponding continuous equation is confirmed, and the torsion-free holonomy is shown to satisfy the equation obtained by varying the holonomy variables. In this sense, the equations of motion obtained in this framework admit a continuum limit that agrees with the field equations of Einstein--Cartan gravity.

These results provide a geometric framework for incorporating torsion into discrete gravity and strengthen the connection between simplicial gravity and the first-order formulation of continuum gravity. Moreover, the close relation between torsion and extrinsic curvature in simplicial geometry \cite{Haggard:2012pm} suggests that the present framework may also provide a useful perspective on the geometric description of extrinsic curvature in discrete gravity.

There are several interesting issues deserving further investigation. It is interesting to study the coupling of the present discrete gravity with matter fields. In particular, it is interesting to see whether the torsion is determined algebraically by the fermions as the case of continuous theory. Also, it is interesting to study the relation between the present discrete gravity and other approaches to quantum gravity, e.g., loop quantum gravity. These issues fare left or future study.

\section*{Acknowledgments}

This work is supported by National Natural Science Foundation of China (NSFC) with Grants No.12275022, No.12675060, No.12505055 and by ``the Fundamental Research Funds for the Central Universities.''

\appendix

\section{Non-Abelian Stokes theorem}\label{sec:non-Abe Stokes theom}

On an $n$-dimensional manifold, let $A$ be a connection (or gauge potential), which is a one-form taking values in a non-Abelian Lie algebra. The non-Abelian curvature (or field-strength) two-form is defined by
\begin{equation}
F(A)={\rm d}A+[A,A].
\end{equation}

\begin{theorem}\label{theom:non-Abe Stokes theom}
The non-Abelian Stokes theorem:
\begin{equation}\label{eq:non-Abe non-Abe Stokes theom}
\mathcal{P}\exp\left(-\oint_{C=\partial S} \!\!\!\! A_a(x) {\rm d}x^a \right)=\mathscr{P}\exp\left(-\iint_S \mathscr{F}\right),
\end{equation}
where $\mathcal{P}$ denotes path ordering and $\mathscr{P}$ denotes an appropriate surface ordering. $\mathscr{F}$, called twisted curvature, is a path-dependent curvature two-form given by the formula
\begin{equation}
\mathscr{F}(x):=h^{-1}_l(O,x) F(x)h_l(O,x),
\end{equation}
where $h_l(O,x)$ is the holonomy of the path $l$ in the surface $S$ joining the base point $O$ of $\partial S$ with the point $x$:
\begin{equation}
h_l(O,x)=\mathcal{P}\exp(-\int_l A_a {\rm d}x^a).
\end{equation}
\end{theorem}

We omit the proof here; it can be found in \cite{Broda:2001id}.

For a Riemann--Cartan manifold, we identify $A$ with the spin connection $\omega^I{}_J$ and $F$ with the curvature two-form $R^I{}_J$. The left-hand side of Eq.~\eqref{eq:non-Abe non-Abe Stokes theom} is then the holonomy $h_C{}^I{}_J$ around the closed curve $C=\partial S$ starting at the point $O$.
We now consider a sufficiently small surface $S$. If the curvature and connection vary smoothly over $S$, the corrections arising from the non-commutativity of twisted-curvature insertions at different points are of higher order than the leading contribution from the surface integral. To leading order, the surface-ordered exponential can therefore be approximated by an ordinary exponential,
\begin{equation}
\mathscr{P}\exp\bigl(\!-\! \iint_S \!\! \mathscr{R}_{ab}{}^I{}_J {\rm d}x^a {\rm d}x^b \!\bigr) \!\approx\! \exp\bigl(\!-\! \iint_S \! \mathscr{R}_{ab}{}^I{}_J {\rm d}x^a {\rm d}x^b\bigr).
\end{equation}
The integral of the twisted curvature can then be approximated as
\begin{equation}\label{eq:non-Abe int scrR-h}
\iint_S \mathscr{R}_{ab}{}^I{}_J{\rm d}x^a {\rm d}x^b \approx -({\rm Ln}h_{C})^I{}_J,
\end{equation}
where the ${\rm Ln}$ denotes the principal branch of the matrix logarithm as discussed in detail in Appendix \ref{sec:Martix Log}.

If we further assume that the holonomy along any path $l$ inside the integration region is sufficiently close to the identity, it can be written as
\begin{equation}
h_l=\mathbf{1}+\delta h_l,\qquad \delta h_l=O(l),
\end{equation}
where $l$ denotes the length of the path and the non-zero matrix elements of $\delta h_l$ are of order $l\ll 1$. Correspondingly,
\begin{equation}
h_l^{-1}=\mathbf{1}-\delta h_l+O(l^2).
\end{equation}
The twisted curvature can then be expanded as
\begin{equation}
\begin{aligned}
\mathscr{R}_{ab}(x)=&h^{-1}_l(O,x) R_{ab}(x)h_l(O,x)\\
\approx &(\mathbf{1}-\delta h_l)R_{ab}(x)(\mathbf{1}+\delta h_l)\\
\approx &R_{ab}(x)+[R_{ab}(x),\delta h_l]+O^2(l).
\end{aligned}
\end{equation}
It follows that the integral of twisted curvature can be approximated by
\begin{equation}\label{eq:non-Abe int scrR-R}
\begin{aligned}
& \iint_S \mathscr{R}_{ab} {\rm d}x^a {\rm d}x^b\\
& \quad \approx \iint_S \left(R_{ab}(x)+[R_{ab}(x),\delta h_l]+O^2(l)\right) {\rm d}x^a {\rm d}x^b.
\end{aligned}
\end{equation}
The twisted curvature depends on the choice of the base point $O$, since the parallel transport $h_l(O,x)$, and hence $\delta h_l$, changes with this choice. In Eq.~\eqref{eq:non-Abe int scrR-R}, however, this dependence enters only through the correction terms involving $\delta h_l$, which are of higher order in the size of the integration region than the leading curvature contribution. Neglecting these higher-order terms, we therefore obtain
\begin{equation}\label{eq:non-Abe int R-h}
\iint_S R_{ab}{}^I{}_J {\rm d}x^a {\rm d}x^b\approx -({\rm Ln}h_C)^I{}_J.
\end{equation}
Although the dependence on the choice of the base point is suppressed at this order, different choices generally lead to different higher-order corrections. To avoid privileging a particular choice, we average over different choices of the base point. While there is no natural discrete set of such choices in a continuum manifold, on a simplicial manifold the base point can naturally be chosen inside each of the simplices surrounding a given hinge. We therefore average the corresponding quantities over these different choices.

\section{Matrix Logarithm}\label{sec:Martix Log}

This appendix collects several results on matrix logarithms used in the main text. We omit the proofs and refer the reader to Refs.~\cite{Haber2018NotesOT, FunctionsOfMatrices}.

A logarithm of a matrix $A\in \mathbb{C}^{n\times n}$ is any matrix $X\in\mathbb{C}^{n\times n}$ such that
\begin{equation}
e^X=A.
\end{equation}
In general, a matrix logarithm, when it exists, is not unique. To characterize matrix logarithms and, in particular, to distinguish a suitable single-valued logarithm, we first recall the definition of a matrix function based on the Jordan canonical form.

Any matrix $A\in\mathbb{C}^{n\times n}$ can be expressed in the Jordan canonical form
\begin{subequations}\label{eq:Matrix-log Jordan form}
\begin{align}
Z^{-1}AZ=J={\rm diag}(J_1,J_2,\dots,J_p),\\
J_k=J_k(\lambda_k)=\begin{bmatrix}
\lambda_{k} & 1 & & \\
& \lambda_{k} & \ddots & \\
& & \ddots & 1 \\
& & & \lambda_{k}
\end{bmatrix}\in \mathbb{C}^{m_k\times m_k},
\end{align}
\end{subequations}
where $Z$ is non-singular and $m_1+m_2+\cdots+m_p=n$. The Jordan matrix $J$ is unique up to the ordering of its Jordan blocks $J_i$, whereas the transformation matrix $Z$ is not unique.

Here, $\lambda_k$ denotes the eigenvalue associated with the $k$-th Jordan block, so different Jordan blocks may have the same value of $\lambda_k$. Let $\mu_1,\dots,\mu_s$ denote the distinct eigenvalues of $A$, namely, the distinct values among $\lambda_1,\dots,\lambda_p$, and let $n_i$ be the order of the largest Jordan block associated with $\mu_i$.

\begin{definition}[Function defined on the spectrum of a matrix]
A function $f$ is said to be defined on the spectrum of $A\in\mathbb{C}^{n\times n}$ if the values
\begin{equation}
f^{(j)}(\mu_i),\ j=0,1,\dots,n_i-1,\ i=1,2,\dots,s,
\end{equation}
exist.
\end{definition}

With this terminology, a matrix function can be defined directly from the Jordan
canonical form.

\begin{definition}[Matrix function via Jordan canonical form]
Let $f$ be defined on the spectrum of $A\in \mathbb{C}^{n\times n}$ and let $A$ have the Jordan canonical form \eqref{eq:Matrix-log Jordan form}. Then
\begin{equation}
f(A) \! :=\! Zf(J)Z^{-1} \!=\! Z{\rm diag}\bigl(f(J_1),f(J_2),\dots,f(J_p)\bigr)Z^{-1},
\end{equation}
where
\begin{equation}\label{eq:Matrix-log f(Jk)}
f(J_k):=\begin{bmatrix}
f\left(\lambda_{k}\right) & f^{\prime}\left(\lambda_{k}\right) & \ldots & \frac{f^{\left(m_{k}-1\right)}\left(\lambda_{k}\right)}{\left(m_{k}-1\right)!} \\
& f\left(\lambda_{k}\right) & \ddots & \vdots \\
& & \ddots & f^{\prime}\left(\lambda_{k}\right) \\
& & & f\left(\lambda_{k}\right)
\end{bmatrix}.
\end{equation}
\end{definition}

For a multi-valued scalar function, such as the logarithm, a choice of branch is required in applying the above definition. In particular, the same branch must be chosen for all Jordan blocks associated with the same eigenvalue in order to obtain a primary matrix function.

Before considering the general form of matrix logarithms, we first recall the condition under which a real matrix admits a real logarithm.

\begin{theorem}[Existence of real logarithm]
A non-singular matrix $A\in \mathbb{R}^{n\times n}$ has a real logarithm if and only if $A$ has an even number of Jordan blocks of each size for every negative eigenvalue.
\end{theorem}

For a non-singular complex matrix, all logarithms can be characterized explicitly
in terms of its Jordan canonical form.

\begin{theorem}[Gantmacher]
Let $A\in \mathbb{C}^{n\times n}$ be non-singular with the Jordan canonical form \eqref{eq:Matrix-log Jordan form}. Then all solutions to $e^X=A$ are given by
\begin{subequations}
\begin{gather}
X=ZU{\rm diag}(L_1^{(j_1)},L_2^{(j_2)},\dots,L_p^{(j_p)})U^{-1}Z^{-1},\\
\shortintertext{where}
L_k^{(j_k)}={\rm Ln}(J_k(\lambda_k))+2j_k\pi i I_{m_k};
\end{gather}
\end{subequations}
${\rm Ln}(J_k(\lambda_k))$ denotes Eq.~\eqref{eq:Matrix-log f(Jk)} with the $f$ taken to be the principal branch of the logarithm, defined by ${\rm Im}({\rm ln}(z))\in(-\pi,\pi]$; $j_k$ is an arbitrary integer; and $U$ is an arbitrary non-singular matrix that commutes with $J$.
\end{theorem}

The Gantmacher theorem shows that a non-singular matrix generally admits multiple logarithms. If the spectrum of the matrix avoids the closed negative real axis $\mathbb{R}^{-}:=(-\infty,0]$, a distinguished logarithm can be selected among them.

\begin{theorem}[Principal logarithm]
Let $A\in \mathbb{C}^{n\times n}$ have no eigenvalues on $\mathbb{R}^-$. There is a unique logarithm $X$ of $A$ all of whose eigenvalues lie in the strip $\{z:-\pi<{\rm Im}(z)<\pi\}$. We refer to $X$ as the principal logarithm of $A$ and write $X={\rm Ln}(A)$.
\end{theorem}

Under a stronger condition on the spectrum of $A$, the principal logarithm admits a convergent Taylor series expansion.

\begin{theorem}[Taylor series expansion of matrix logarithm]\label{theom:Matrix-log ln(A)-Taylor series}
Let $A\in\mathbb{C}^{n\times n}$ satisfy
\begin{equation}
\rho(A-I)<1,
\end{equation}
where $\rho$ denotes the spectral radius. Then the principal logarithm of $A$ admits the convergent expansion
\begin{equation}\label{eq:Matrix-log ln(A)-Taylor series}
{\rm Ln}A=\sum_{n=1}^{\infty}\frac{(-1)^{n+1}}{n}(A-I)^n.
\end{equation}
\end{theorem}

Under the same condition, varying Eq.~\eqref{eq:Matrix-log ln(A)-Taylor series} term by term gives
\begin{equation}\label{eq:Matrix-log del(lnA)-Taylor series}
\delta({\rm Ln}A)=\sum_{n=1}^\infty\sum_{a=0}^{n-1} \frac{(-1)^{n+1}}{n}(A-I)^a(\delta A)(A-I)^{n-a-1}.
\end{equation}

Finally, the principal logarithm is covariant under similarity transformations.

\begin{theorem}[Similarity transformation of the principal logarithm]\label{theom:Matrix-log ln(SAS^-1)}
Let $A\in\mathbb{C}^{n\times n}$ have no eigenvalues on $\mathbb{R}^-$, and let $S\in\mathbb{C}^{n\times n}$ be non-singular. Then
\begin{equation}
{\rm Ln}(SAS^{-1})=S({\rm Ln}A)S^{-1}.
\end{equation}
\end{theorem}

\bibliographystyle{ieeetr}
\bibliography{ref}

\end{document}